\documentclass[mathpazo]{cicp}

\usepackage{graphicx}
\usepackage{amsmath}
\newcounter{pamataj}

\begin{document}
%%%%% title : short title may not be used but TITLE is required.
% \title{TITLE}
% \title[short title]{TITLE}
\title[short title]{Relativistic quantum-statistical distributions}
\title{Generation of random deviates for relativistic quantum-statistical distributions}

%%%%% author(s) :
% single author:
% \author[name in running head]{AUTHOR\corrauth}
% [name in running head] is NOT OPTIONAL, it is a MUST.
% Use \corrauth to indicate the corresponding author.
% Use \email to provide email address of author.
% \footnote and \thanks are not used in the heading section.
% Another acknowlegments/support of grants, state in Acknowledgments section
% \section*{Acknowledgments}
%\author[O.~Author]{Only Author\corrauth}
%\address{School of Mathematical Sciences, Beijing Normal University,
%Beijing 100875, P.R. China}
%\email{{\tt author@email} (O.~Author)}

% multiple authors:
% Note the use of \affil and \affilnum to link names and addresses.
% The author for correspondence is marked by \corrauth.
% use \emails to provide email addresses of authors
% e.g. below example has 3 authors, first author is also the corresponding
%      author, author 1 and 3 having the same address.
\author[B.~Tom\'a\v{s}ik, I.~Melo, J. Cimerman]%
      {Boris Tom\'a\v{s}ik\affil{1}\comma\affil{2}\comma\corrauth,
       Ivan Melo\affil{1}\comma\affil{3},
	Jakub Cimerman\affil{4}
}
\address{\affilnum{1}\ Univerzita Mateja Bela,
            Tajovsk\'eho 40,
           97401 Bansk\'a Bystrica, Slovakia \\
            \affilnum{2}\ FNSPE,
            Czech Technical University in Prague, B\v{r}ehov\'a 7, 11519 Prague 1,\\
           Czech Republic\\
	     \affilnum{3}\ \v{Z}ilinsk\'a univerzita, Univerzitn\'a 1, 01001 \v{Z}ilina, Slovakia\\
	     \affilnum{4}\ FMFI, Comenius University, Mlynsk\'a dolina, 84248 Bratislava, Slovakia
}
 \email{{\tt boris.tomasik@cern.ch} (B. Tom\'a\v{s}ik)}
%          {\tt zhao@email} (A.~Zhao)}
% \footnote and \thanks are not used in the heading section.
% Another acknowlegments/support of grants, state in Acknowledgments section
% \section*{Acknowledgments}

%%%%% Begin Abstract %%%%%%%%%%%
\begin{abstract}
We provide an algorithm for generation of momenta (or energies) of relativistic particles 
according to the relativistic Bose-Einstein or Fermi-Dirac distributions. The algorithm uses 
rejection method with effectively selected comparison function so that the acceptance rate 
of the generated values is always better than 0.9. It might find its use in Monte-Carlo
generators of particles from reactions in high-energy physics.
\end{abstract}
%%%%% end %%%%%%%%%%%

%%%%% AMS/PACs/Keywords %%%%%%%%%%%
%\pac{}

%%%% maketitle %%%%%
\maketitle

%%%% Start %%%%%%
\section{Motivation}
\label{s:motiv}

In projects related to multiparticle production in hadronic or nuclear collisions it is often 
demanded to generate a large number of particles with momenta distributed according to 
relativistic Bose-Einstein of Fermi-Dirac distribution. Here one has to take into account 
the total energy (i.e.\ including the mass) when evaluating the exponent of the distributions
\begin{equation}
f(\vec p) = \frac{1}{(2\pi \hbar)^3} \left [
\exp\left (\frac{c\sqrt{m^2c^2 + p^2} - \mu}{k_B T}
\right ) + q
\right ]^{-1}
\end{equation}
where $m$ is the mass of the particles, $p = |\vec p|$, 
$\mu$ is the chemical potential and $T$ is temperature. 
Parameter $q$ assumes the value of 1 for fermions and $-1$ for bosons. 
For the generation of invariant momentum distributions one also needs this distribution 
multiplied with the energy
\begin{equation}
S(\vec p) = \frac{c\sqrt{m^2c^2 + p^2}}{(2\pi \hbar)^3} \left [ 
\exp\left (\frac{c\sqrt{m^2c^2 + p^2} - \mu}{k_B T}
\right ) + q
\right ]^{-1}
\end{equation}
As the distribution is spherically symmetric, the angles are trivially integrated  and we 
are left with the distributions for the size of the momentum vector. In order to make it suitable 
for a general procedure, it is expressed  with the help of  dimensionless variable 
\begin{equation}
x = \frac{p}{mc}\,  .
\end{equation}
Thus we get 
\begin{equation}
\label{e:dist}
f(x) =  x^2 \, 
\left [
\exp\left ( A (\sqrt{1 + x^2} - M )
\right ) + q
\right ]^{-1}
\end{equation}
or for the other distribution
\begin{equation}
\label{e:ldist}
S(x) = \sqrt{1+x^2}\, x^2 \, 
\left [
\exp\left ( A (\sqrt{1 + x^2} - M )
\right ) + q
\right ]^{-1}\, ,
\end{equation}
where
\begin{equation}
A = \frac{mc^2}{k_BT}\, , \qquad M = \frac{\mu}{mc^2}\,  .
\end{equation}
In these functions we have suppressed the constant pre-factors which also  contain 
dimensions.

For the Monte Carlo generation we shall proceed with the dimensionless distributions without 
the dimensionfull pre-factors. 

A similar algorithm for the generation of relativistic Maxwellian distribution has been 
reported in \cite{Swisdak,Devroye}. 
In our work we properly account for quantum statistics and 
allow for non-zero chemical potential, which can influence the momentum distribution. 
We describe the procedure for the distribution (\ref{e:ldist}) with the energy factor. 
The procedure for the distribution (\ref{e:dist}) can easily be derived along the same 
steps as we shall proceed.

%%%%%%%%%%%%%%%%%%%%%%%%%%%%%%%%%%%%%%%%
\section{The algorithm}
\label{s:algo}

We demonstrate in the Apendix that the distribution  (\ref{e:ldist}) is  
log-concave for large enough $x$, i.e.\
its logarithm is a concave function. For such a distribution 
there always exists an exponential that 
is everywhere above the demanded distribution. One can generate random 
deviates according to the exponential and use the rejection method\cite{Devroye}. 

\begin{figure}[t]
\centerline{\includegraphics[width=0.7\textwidth]{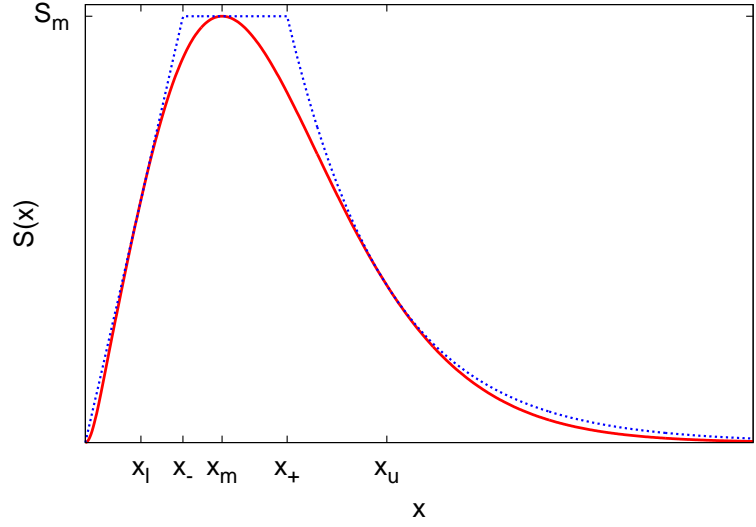}}
\caption{\label{f:schema}%
Dimensionless Bose-Einstein distribution with energy pre-factor according to 
Eq.~(\ref{e:ldist}). The shape corresponds to $A=2/3$ and $M=3/4$ but we 
have suppressed the values on the axes in order to demonstrate the comparison function
and locate the important points for the determination of the comparison 
function. 
}
\end{figure}

In order to achieve smallest possible rejection rate we use piecewise analytic 
comparison function, as indicated in Fig.~\ref{f:schema}. 
%%%%%%%%%%%%%%%%%%%%%%%%%%%%%%%%%%%%%%%%%%%%%%%%%%%%
%%%%%%%%%%%%%%%%%%%%%%%%%%%%%%%%%%%%%%%%%%%%%%%%%%%%
The three pieces are determined so that 
\begin{itemize}
\item for $x \le x_-$ the comparison function is linear;
\item for $x_- < x \le x_+$ the comparison function is constant and equal to the value 
of the distribution at the mode;
\item for $x > x_+$ the comparison function is exponential. 
\end{itemize}
The joint points $x_-$ and $x_+$ are chosen so that the comparison function 
is always continuous. 

For the determination of the comparison function we thus need to determine the five
points indicated on the horizontal axis.
\begin{description}
\item{$x_m$} The mode of $S(x)$. This is obtained easily by differentiating and we get 
\begin{equation}
2 + 3x_m^2 - \frac{A x_m^2 \sqrt{1+x_m^2}}{1 + q\exp\left ( -A(\sqrt{1+x_m^2}-M)\right )}
=0\,  .
\label{e:xm}
\end{equation}
Unfortunately this expression cannot be solved analytically and numerical methods must 
be invoked. 

Subsequently, the value of $S(x)$ at the mode can be determined
\begin{equation}
S_m = S(x_m) = \sqrt{1+x_m^2}\, x_m^2 \, 
\left [
\exp\left ( A (\sqrt{1 + x_m^2} - M )
\right ) + q
\right ]^{-1}\, .
\label{e:Sm}
\end{equation}
%%%%%%%%%%%%%%%%%%%%
\item{$x_l$} The point left from the mode in which the linear comparison function touches 
the distribution. It is found from the condition for the derivative of the distribution
\[
\left . \frac{d S(x)}{dx}\right |_{x = x_l} = 
\frac{S(x_l)}{x_l} 
\]
which leads to 
\begin{equation}
1 + 2x_l^2 - 
\frac{A x_l^2 \sqrt{1+x_l^2}}{1 + q\exp\left ( -A(\sqrt{1+x_l^2}-M)\right )}
= 0 \,  .
\label{e:xl}
\end{equation}
The slope of the liner comparison function is then 
\begin{equation}
K = \frac{S(x_l)}{x_l} = 
\frac{x_l \sqrt{1+x_l^2}}{\exp\left ( A ( \sqrt{1+x_l^2} - M) \right ) + q}\,   .
\end{equation}
%%%%%%%%%%%%%%%%%%%%%%%
\item{$x_-$} The point in which the linear part of  the comparison function and its 
constant part meet. It can be determined as 
\begin{equation}
x_- = \frac{S_m}{x_l\sqrt{1+x_l^2}} \left [
\exp\left ( 
A(\sqrt{1+x_l^2} - M)
\right ) + q
\right ]\,  .
\label{e:xmin}
\end{equation}
The knowledge of $x_-$ also allows to express 
\begin{equation}
K = \frac{S_m}{x_-}\,  .
\end{equation}
%%%%%%%%%%%%%%%%%%%%%%%%
\item{$x_u$} The point in which the exponential part of the comparison function touches
the distribution. 
Above the mode the distribution is log-concave. Therefore, $x_u$ 
can be chosen anywhere above
$x_m$. However, we checked that,
the acceptance rate is optimised with $x_u$ chosen so that the 
distribution there drops to  $1/e$ of its maximum value. We get the value by solving 
the equation for the logarithms of the distribution
\[
\ln S(x_u) = \ln S_m - 1\,  ,
\]
which leads to 
\begin{equation}
1 - \ln S_m + \frac12 \ln (1+x_u^2) + 2\ln x_u -
\ln \left [
\exp \left (
A (\sqrt{1+x_u^2} - M)
\right ) + q
\right ] = 0\,  .
\label{e:xu}
\end{equation}
Again, this equation must be solved numerically. 

Once $x_u$ is determined, we can determine the slope parameter of the exponential 
comparison function. It is given by the logarithm of the distribution. Thus 
\begin{equation}
\lambda = -\left . \frac{d\ln S(x)}{dx}\right |_{x=x_u}
= \frac{x_u}{\sqrt{1+x_u^2}} \left [
1 - A \frac{\exp\left ( A(\sqrt{1+x_u^2} - M)  \right )}%
{\exp\left ( A(\sqrt{1+x_u^2} - M )   \right ) + q}
\right ] - \frac{2}{x_u}\, .
\label{e:lambda}
\end{equation}
%%%%%%%%%%%%%%%%%%%%%%%%%%%%%%%%%%%%%%%%
\item{$x_+$} Finally, this is the point in which the constant part of the comparison 
function and its exponential part join. It is determined from a simple equation 
\begin{equation}
x_+ = x_u - \frac{1}{\lambda}\,  .
\label{e:xplus}
\end{equation}

Once we have $x_+$, we also know the exponential part of the comparison function 
which reads
\begin{equation}
S'(x) = S_m e^{-\lambda(x - x_+)}\,  .
\end{equation}
\end{description}

%%%%%%%%%%%%%%%%%%%%%%%%%%%%%%%%%%%%%%%%%%%

Thus we can formulate the comparison function
\begin{equation}
S'(x) = \left \{
\begin{array}{lcl}
\frac{S_m}{x_-} x & : & x \le x_-\\
S_m & : &  x_- < x \le x_+ \\
S_m e^{-\lambda (x - x_+)} & : & x > x_+
\end{array}
\right .
\end{equation}

In order to use this comparison function as probability density (after normalisation) for 
random variate generation we need the values
\begin{subequations}
\label{e:ys}
\begin{eqnarray}
y_- & = & \int_0^{x_-} S'(x) dx = \frac{1}{2} S_m x_- \\
y_+ & = & \int_0^{x_+} S'(x) dx = S_m\left (x_+ - \frac{1}{2} x_-\right )\\
y_\infty & = & \int_0^\infty S'(x) dx = 
S_m\left (x_+ - \frac{1}{2} x_- + \frac{1}{\lambda} \right )\,  .
\end{eqnarray}
\end{subequations}
The inverse if the integral of $S'(x)$ is
\begin{equation}
\label{e:inverse}
x(y) = \left \{
\begin{array}{lclcl}
\sqrt{\frac{2 x_- y}{S_m}} & : & y \le y_- & \mathrm{gives} & x \le x_-\\
\frac{1}{2} x_- + \frac{y}{S_m} & : & y_- < y \le y_+ & \mathrm{gives} & x_- < x \le x_+\\
x_+ - \frac{1}{\lambda}\ln\left [
1 + \left ( x_+ - \frac{1}{2}x_-\right ) \lambda - \frac{\lambda}{S_m} y 
\right ] &
: & y_+ < y \le y_\infty & \mathrm{gives} & x > x_+
\end{array}
\right . 
\end{equation}

For the rejection step we need the probabilities to accept the generated value of 
$x$. They are given as ${\cal P}(x) = S(x)/S'(x)$. In the three intervals they read
\begin{subequations}
\label{e:accprobs}
\begin{align}
{\cal P}(x)  & = \frac{x_- x \sqrt{1+x^2}}{%
S_m\left [ \exp \left (A(\sqrt{1+x^2} - M)\right ) + q\right ]} & : &&& x\le x_-\\
{\cal P}(x)  & = \frac{x^2 \sqrt{1+x^2}}{%
S_m\left [ \exp \left (A(\sqrt{1+x^2} - M)\right ) + q\right ]} & : &&& x_- < x \le x_+\\
{\cal P}(x)  & = \frac{x^2 \sqrt{1+x^2}\exp\left ( \lambda (x - x_+)\right )}{%
S_m\left [ \exp \left (A(\sqrt{1+x^2} - M)\right ) + q\right ]} & : &&& x_+ < x
\end{align}
\end{subequations}

%%%%%%%
Now we have collected all expressions needed to build up the algorithm. Because of the
need to numerically solve a few equations the procedure may be lengthy if one needs
to generate just one random value. However, if many values must be generated for the 
same temperature, chemical potential and particle mass, then all parameters can 
be calculated first and then used repeatedly. Thus the first part of the algorithm 
is the calculation of the parameters:
\begin{enumerate}
\item Determine $x_m$ by solving Eq.~(\ref{e:xm}).
\item Calculate $S_m$ from 
Eq.~(\ref{e:Sm}). 
\item Determine $x_l$ by solving Eq.~(\ref{e:xl}).
\item Calculate $x_-$ from Eq.~(\ref{e:xmin}).
\item Determine $x_u$ by solving Eq~(\ref{e:xu}).
\item Calculate $\lambda$ from Eq.~(\ref{e:lambda}).
\item Calculate $x_+$ from Eq.~(\ref{e:xplus}).
\item For later convenience calculate also the values of $y_-$, $y_+$, and $y_\infty$ 
from Eqs.~(\ref{e:ys}).
\setcounter{pamataj}{\value{enumi}}
\end{enumerate}
This is the common part of the preparation. Then, in order to generate a value, 
follow these steps: 
\begin{enumerate}
\setcounter{enumi}{\value{pamataj}}
\stepcounter{pamataj}
\item Generate uniform random deviate $y$ from the interval $[0,y_\infty]$.
\item Calculate $x = x(y)$ from Eq.~(\ref{e:inverse}).
\item Accept the value of $x$ with the probability given by Eqs.~(\ref{e:accprobs}). 
If the value is not accepted, return to step \arabic{pamataj}.
\end{enumerate}

%%%%%%%%%%%%%%%%%%%%%%%%%%%%%%%%%%%%%%
\section{Illustration of results}
\label{s:illus}

\begin{figure}[t]
\centerline{%
\includegraphics[width=0.49\textwidth]{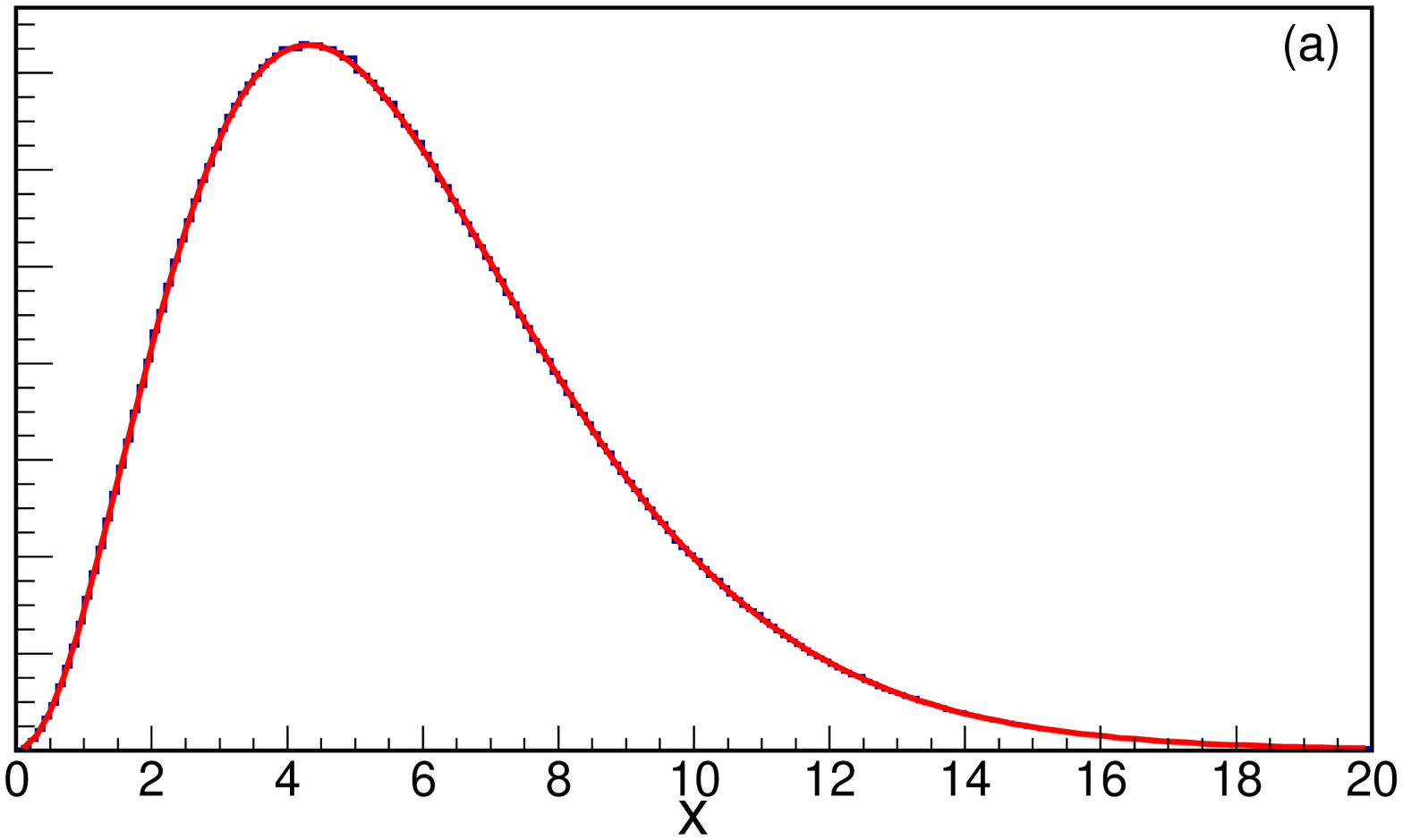}
\includegraphics[width=0.49\textwidth]{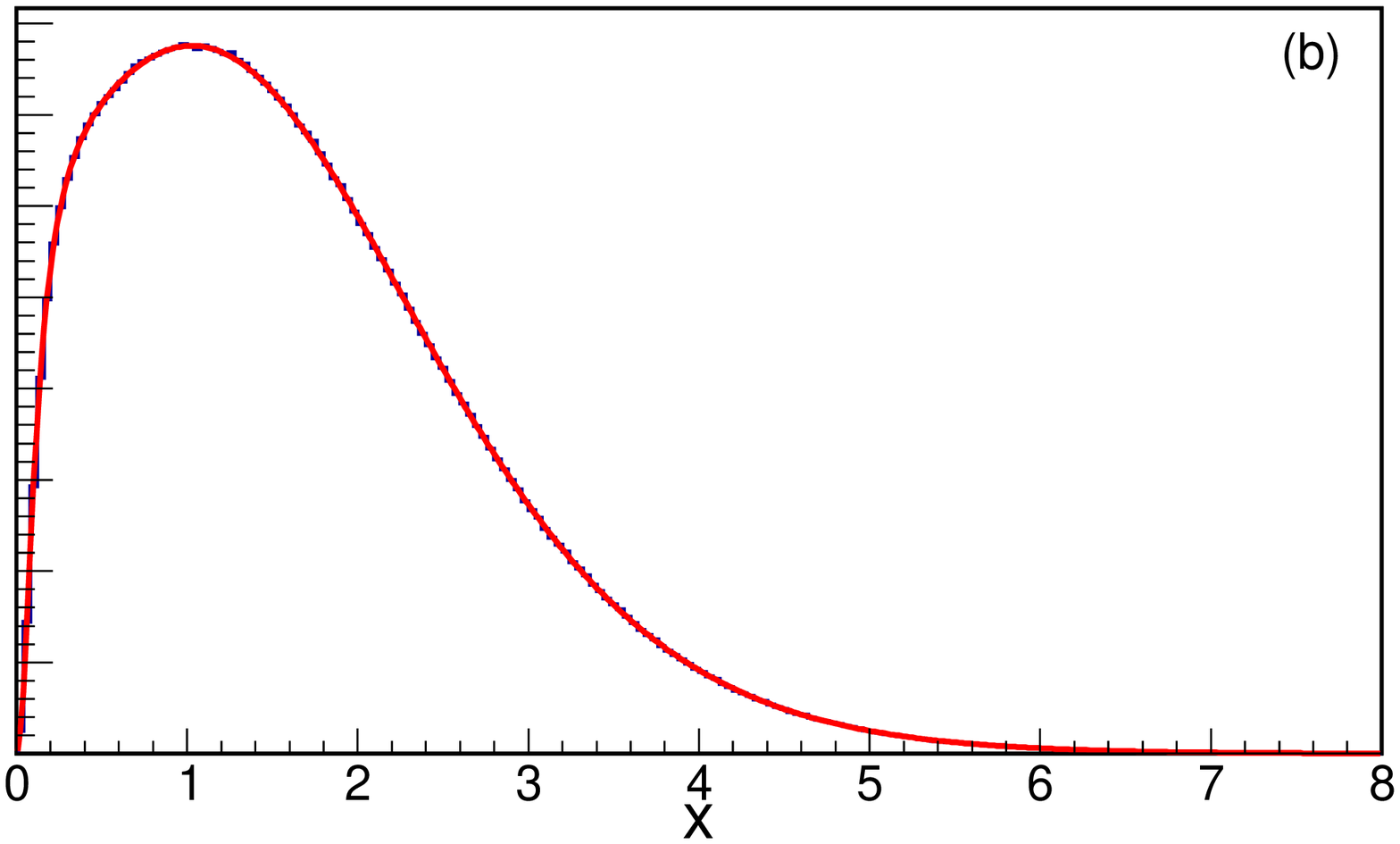}
}
\centerline{%
\includegraphics[width=0.49\textwidth]{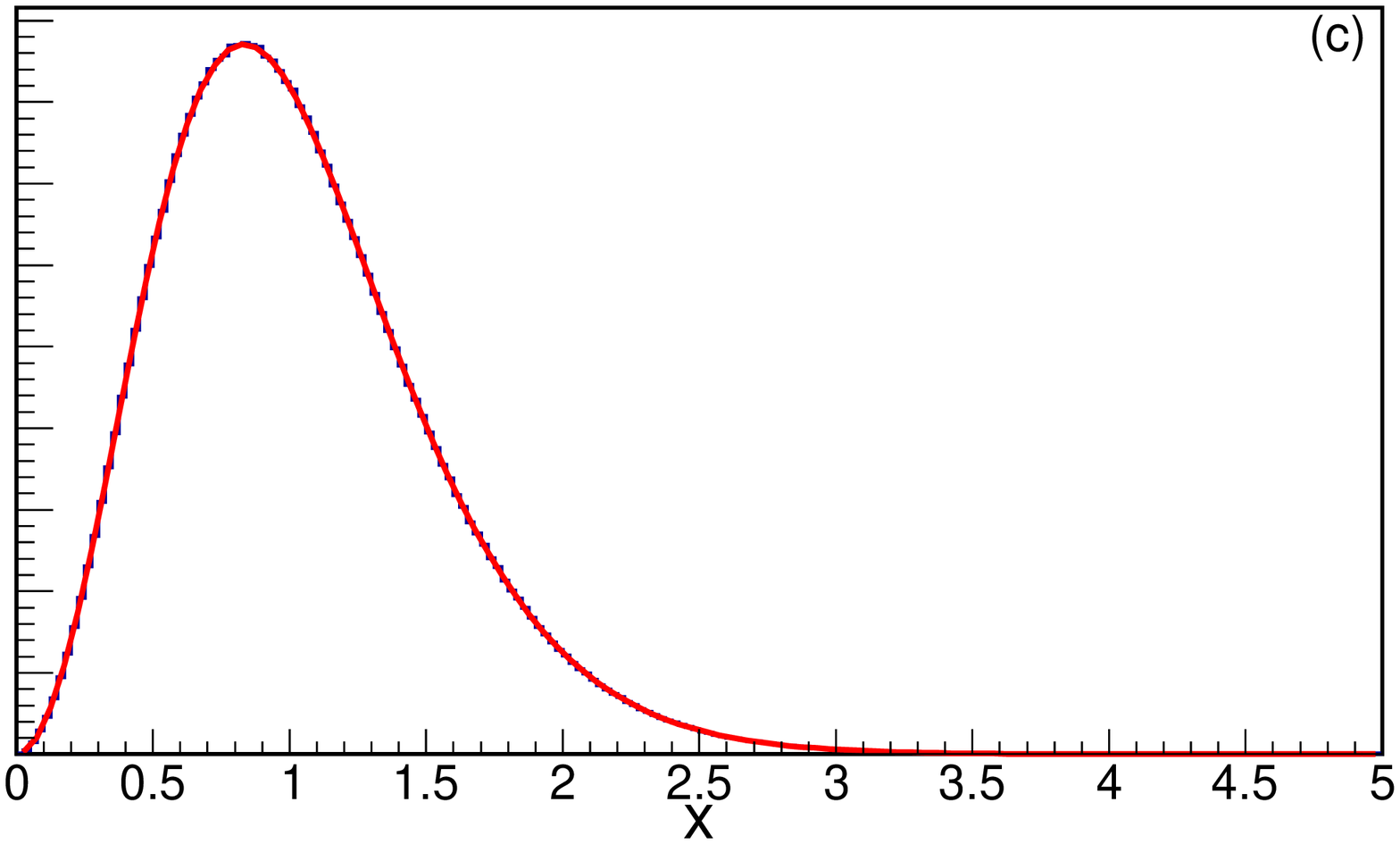}
\includegraphics[width=0.49\textwidth]{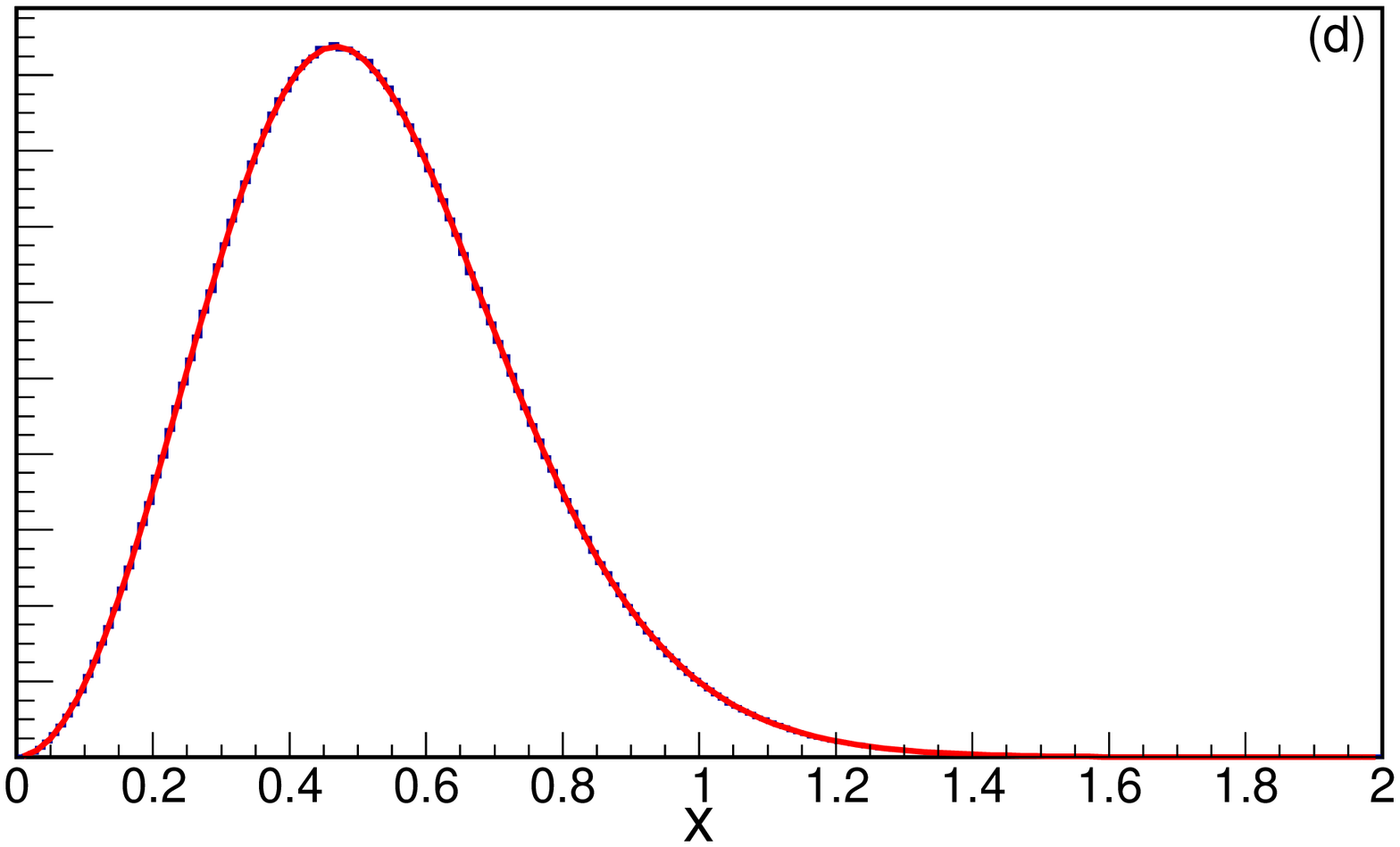}
}
\caption{\label{f:histos}%
Histograms of $10^7$ random deviates fitted by function $S(x)$ with only the 
absolute normalisation as a fit parameter. Values of $A$ and $M$ are fixed in the
fit function in order to be the same as in the Monte Carlo generation. The values 
are: 
(a) $A = 2/3$, $M =  0$, bosons 
(e.g.\ pions with $\mu = 0$ at temperature $k_B T = 207$~MeV);
(b) $A = 2$, $M =  0.993$, bosons 
(e.g.\ pions with $\mu = 137$~MeV$/c^2$ 
at temperature $k_B T = 69$~MeV, close to condensation);
(c) $A = 4.536$, $M =  0$, fermions 
(e.g.\ protons with $\mu = 0$ at temperature $k_B T = (3/2) m_\pi c^2 = 207$~MeV);
(d) $A = 13.609$, $M =  0.9989$, fermions 
(e.g.\ protons with $\mu = 938$~MeV at temperature $k_B T =  m_\pi c^2/2 = 69$~MeV).
}
\end{figure}

We have tested this algorithm in wide range of parameters $A$ and $M$. Particle 
masses were chosen both smaller than temperature so that large momenta 
are available and also much larger than temperature so that the momenta are 
practically non-relativistic. Chemical potentials up to the value of particle 
mass for bosons, i.e.~the point of condensation, were tested, as well
(Figure~\ref{f:histos}). In all cases
the acceptance rate was around 90\%. This shows that the comparison 
function is very well adapted to the present problem. 
%%%%%%%%%%%%%%%%%%%%%%%%%%%%%%%%%%%%%%%%%%%%%%%%%%%%
%%%%%%%%%%%%%%%%%%%%%%%%%%%%%%%%%%%%%%%%%%%%%%%%%%%%

\section{Conclusions}
\label{s:conc}

The presented algorithm has been successfully implemented in an upgrade of the 
Monte Carlo event generator DRAGON \cite{dragon}, which serves for the generation of hadrons 
produced in high energy nuclear collisions. It is, however, general and can serve in 
any other application where relativistic momenta must be generated from 
quantum-statistical distributions. 

%%%%%%%%%%%%%%%%%%%%%%%%%%%%%%%%%%%%%%%%%%%%%%%%%%%%%%%%%

%%%% Acknowledgments %%%%%%%%
\section*{Acknowledgments}

We gratefully acknowledge financial support by grants 
APVV-0050-11, VEGA 1/0469/15 (Slovakia) and 
M\v{S}MT grant  LG13031 (Czech Republic). 
The reported algorithm has been used in our software which run in 
the High Performance Computing Center of the Matej Bel University in Bansk\'a Bystrica 
using the HPC infrastructure acquired in project ITMS 26230120002 and 26210120002 
(Slovak infrastructure for high-performance computing) supported by the Research \& Development Operational Programme funded by the ERDF.

%%%%%%%%%%%%%%%%%%%%%%%%%%%%%%%%%%%%%%%%%%%%%%%%%

\appendix

\section{Log-concave distribution}

In this Appendix we demonstrate that the distribution $S(x)$ according 
to Eq.~(\ref{e:ldist}) is indeed log-concave on the interval above the mode. 
Therefore, an exponential function which touches $S(x)$ from above in one point 
will never be smaller than $S(x)$. 

The calculation is straightforward. We take the second derivative of 
$\ln S(x)$. For fermions $(q = 1)$, this leads to 
\begin{equation}
\frac{d^2 \ln S(x)}{dx^2} = - \frac{1}{1+x^2} 
\left \{
\frac{2}{x^2} + 1 + \frac{2x^2}{1+x^2} + \frac{A}{\sqrt{1+x^2}} \left ( 1 - s_f(x) \right )
+A^2 x^2 s_f(x) \left ( 1 - s_f(x) \right )
\right \}
\label{e:fder}
\end{equation}
where 
\[
s_f(x) = \frac{1}{\exp\left [ A ( \sqrt{1+x^2} - M) \right ] + 1}\,  .
\]
Note that $s_f(x)\le 1 $ for any $x$. Therefore, $(1-s_f(x))\ge 0$ and all terms in the bracket 
in Eq.~(\ref{e:fder}) are non-negative. In summary, we see that \emph{for fermions}
\begin{equation}
\frac{d^2 \ln S(x)}{dx^2} < 0
\end{equation}
and thus the distribution is log-concave everywhere. 

The case of bosons is slightly more involved. Again, we take the second derivative
\begin{equation}
\frac{d^2\ln S(x)}{dx^2} = -\frac{1}{1+x^2} \left \{
\frac{2}{x^2} + 1 + \frac{2x^2}{1+x^2} + \frac{A}{\sqrt{1+x^2}} \left ( 1 + s_b(x) \right )
- A^2 x^2 s_b(x) \left ( 1 + s_b(x) \right ) 
\right \}
\label{e:bder}
\end{equation}
where 
\[
s_b(x) = \frac{1}{\exp\left [  A (\sqrt{1+x^2} - M )\right ] - 1}\,  .
\]
Note  the change of the sign in $(1+s_b(x))$ and in front of the last term. Due to this, 
for bosons the second derivative may become positive in some cases. We want to 
demonstrate that such pathological intervals are always below the mode of $S(x)$. 

For $x\to \infty$ the terms $s_b(x)$ go to 0 exponentially (we chose the letter $s$ for 
``small''), and one can inspect that 
\[
\lim_{x\to \infty} \frac{d^2\ln S(x)}{dx^2}  =  0\,   
\]
and the value of the limit is being approached from below.  
Thus the second derivative is either negative everywhere or there is a point $x=x_c$ where 
it crosses the horizontal axis and stays negative for $x>x_c$. It is enough to show that 
$x_c < x_m$.

\begin{figure}[t]
\centerline{\includegraphics[width=0.84\textwidth]{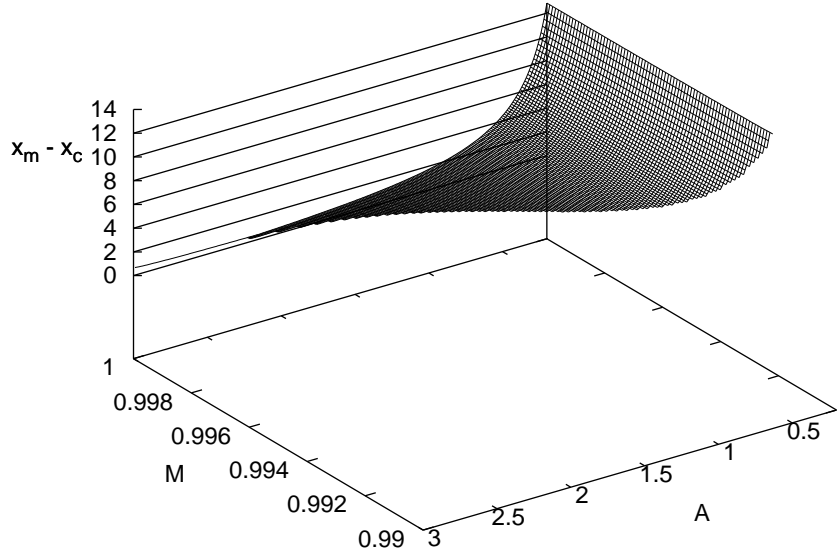}}
\caption{\label{f:xcxm}%
The difference in positions $x_m - x_c$ as function of $A$ and $M$. 
Values are not plotted if the second derivative $d^2\ln S(x)/dx^2$ 
for bosons stays negative everywhere. One sees that always $x_c < x_m$.
}
\end{figure}

It turns out that the second derivative becomes positive only if the particles are light 
with $A\le 3$ and $M$ is  close to 1, which is quite an extreme case. 
(Recall that for bosons $M$ must be smaller than 1 and 
$M=1$ corresponds to the condensation point where the distribution does no longer apply.) 
In such a case, for small values of $x$ one obtains $s_b \gg  1$ and 
the last term in Eq.~(\ref{e:bder}) prevails. We have scanned the whole relevant parameter 
region of $A$ and $M$ and checked that always $x_c < x_m$ (Figure~\ref{f:xcxm}).
%%%%%%%%%%%%%%%%%%%%%%%%%%%%%%%%%%%%%%%%%%%%%%%%%%%%
%%%%%%%%%%%%%%%%%%%%%%%%%%%%%%%%%%%%%%%%%%%%%%%%%%%%
In all other cases the function $S(x)$ is log-concave everywhere. 

We conclude that it is safe to use the exponential comparison function in the 
interval $(x_m,\infty)$.

%%%% Bibliography  %%%%%%%%%%

\end{document}